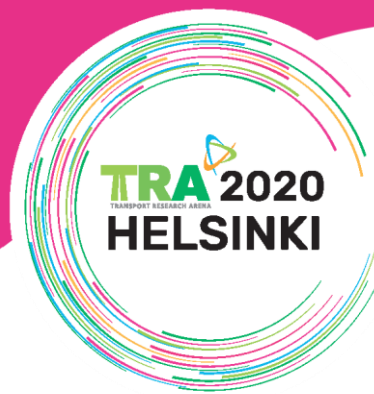

# SPRINT: Semantics for PerfoRmant and scalable INteroperability of multimodal Transport

Mersedeh Sadeghi[a], Petr Buchníček [b], Alessio Carenini[c], Oscar Corcho[d], Stefanos Gogos[e]*, Matteo Rossi[a], Riccardo Santoro[f],

[a]*Politecnico di Milano, Dipartimento di Elettronica, Informazione e Bioingegneria, Piazza L. Da Vinci 32, 20133, Milan, Italy*

[b] *OLTIS Group, Dr Milady Horakove 1200/27a, 77900, Olomouc, Czech Republic*

[c]*CEFRIEL, Viale Sarca, 226, 20126 Milano, Italy*

[d] *Ontology Engineering Group, Departamento de Inteligencia Artificial, Facultad de Informática, Universidad Politécnica de Madrid, Campus de Montegancedo s/n 28660, Boadilla del Monte, Spain*

[e]*UNIFE (European Rail Industry), 221 Avenue Louise, Brussels 1050, Belgium*

[f] *TRENITALIA, 1 Piazza della Croce Rossa, 00161 Roma, Italy*

**Abstract**

Interoperability is a long-standing challenge slowing down the digitalization of mobility systems and the provision of full mobility-as-a-service offerings. This paper presents early results achieved by the SPRINT project (Semantics for PerfoRmant and scalable INteroperability of multimodal Transport), an integral part of the Shift2Rail IP4 work programme, in meeting the challenge. It introduces the conceptual architecture and describes the demonstrator implementation of the Interoperability Framework (IF), a solution designed to support servitization in the mobility domain in two directions: From the Travel Application perspective, the IF provides semantically-consistent abstractions of distributed ICT resources offered by Service Provider Companies, insulating them from the "mechanics" of operating remotely over networks and across multiple communication protocols and/or data formats; from the Service Provider Companies perspective, the IF allows leveraging their native ICT computing environment and resources as elements of an end-to-end integrated intermodal mobility solution, insulating them from the specifics of the customer front-end applications.

---

\* Corresponding author. Tel.: + 32 (0) 2 431 04 62;

*E-mail address:* stefanos.gogos@unife.org

1. Introduction

The provision of truly customer-centric mobility services across the Single European Transport Area requires advanced ICT applications that can discover, access and combine mobility solutions from multiple Service Provider Companies. From a digital system engineering point of view, these applications must be able to coordinate the execution of complex computational tasks that are inherently distributed on multiple heterogeneous "nodes" of an open network with no centralized control.

"Distribution transparency", as defined by the "ITU-T Rec. X.903 | ISO/IEC 10746-3: Architecture standard for open distributed processing" (ISO/IEC 10746-3:2009, 2009) is a key concept for the design and implementation of such an interoperable distributed computing ecosystem. Consequently, the main design goals of the Shift2Rail IF consist of:

- Access transparency, which masks differences of data representation and invocation mechanisms for services between systems.
- Location transparency, which masks the need for an application to have information about location in order to invoke a service.
- Relocation transparency, which masks the relocation of a service from applications using it.
- Replication transparency, which masks the fact that multiple copies of a service may be provided in order to ensure reliability and availability.

Access and location transparencies are achieved by leveraging semantic interoperability principles and technologies: knowledge about the domain problem, which is typically held by human analysts and programmers, is formalized in a set of logical statements, or "axioms", written in a standard computer language available for machine processing. Human knowledge is thus transferred to machines and shared by them. Any particular representation of concepts and relationships in a specific data structure is associated, through a process of annotation, with its interpretation in terms of the domain problem. Machines can, therefore, discover and leverage equivalence relationship between different data formats with a common meaning, and automate, therefore, the translation across these formats. Automated computer logical inference replaces human programming of software to operate on different – but equivalent – data formats however they may be exchanged. Semantically-annotated data are furthermore linked to constitute a shared semantic graph, or "web of transport data", whose physical distribution across networked machines is invisible to consuming services.

The term "servitization" was introduced by S. Vandermerwe and J. Rada in 1988 to indicate a "movement" whereby "modern corporations are increasingly offering fuller market packages or 'bundles' of customer-focused combinations of goods, services, support, self-service and knowledge [..] with services beginning to dominate" (Vandermerwe & Rada, 1998). Since then a vast scientific and technical literature on the concept has developed and numerous case studies have been published extending beyond the now classical Xerox Corp. 'photocopier' and Rolls Royce 'power-by-the-hour' benchmarks[†]. While different characterizations of the concept exist, for the purpose of this paper we stress the following features:

- The customer buys the outcome of, or capability enabled by the product, not the product itself, e.g. document management rather than printers and photocopiers, airplane thrust rather than jet engines.
- The Customer pays for use and assured performance on the basis of relatively long-term tailored contracts
- The Customer is actively engaged in the definition of the services 'bundle' and in the services' usage and performance monitoring

These three characteristics are common, albeit in various degrees of sophistication, to all use cases. As a consequence, in all use cases the successful delivering company is equipped with an accurate service performance monitoring and control capability, including for usage metering, that extends to individual remote device installations and customers. We may note that 'servitization', both the thing and the term, were introduced in business practice and in the literature long before the thing and the term 'digitalization', and in fact before the widespread availability of the internet and the world wide web for commercial use. This observation should help as a reminder that, on one hand, digitalization does not, in and of itself, produce servitization, and that, on the other hand, a servitization strategy should shape the digital transformation rather than the opposite. As an enabler of servitization, the essential role of digitalization may be summarized as:

- Bringing many more products that deliver customer-centric 'outcomes', and bringing many more customers themselves (through their own devices), in the scope of vastly enhanced service metering, performance and control monitoring systems, thus enormously multiplying the range of valuable 'things' that can be delivered 'as-a-service'.
- Enabling such systems to operate and adjust at least in part automatically in real-time.
- Integrating multiple 'as-a-service' offerings across different infrastructures and companies.
- Providing customers themselves with the ability to compose, and/or tailor as-a-service bundles from multiple infrastructures and companies, and to interact with them.

---

[†] See for example: (Lay, 2014) discussing servitization in nine different sectors including the photocopier and aircraft industries.

Mobility-as-a-service is indeed an extension to mobility of the 'as-a-service' concept made possible by digitalization: in the case of a free-floating car sharing service, for example, cars and customers (through their own devices) are now computing devices that participate in an application that measures, monitors and controls the service delivery adjusting it to specific conditions and/or customer input, where the customer pays for usage of the outcome, i.e. mobility, and is not concerned with the ownership or servicing of the vehicle. But the same example also offers an additional and critical insight into the nature, or 'global architecture', of such an application: it is *distributed* over multiple independent *interoperable* and *heterogeneous* network infrastructures, with application-level 'intelligence' located at the *edge* of, *not* embedded in the infrastructures themselves. A car sharing mobility-as-a-service "application" sits, in fact, at "the edge" of the city road and traffic management, the internet, telecommunications, satellite GPS positioning, credit/financial circuits and, in the case of electric vehicles, electric power grid infrastructures, *none of which* were ever designed for that particular application-level functional or business requirements. This 'global architecture' insight is known in the industry as the "end-to-end principle" or "argument":

> "End-to-end arguments are a kind of "Occam's razor" when it comes to choosing the functions to be provided in a communication subsystem. Because the communication subsystem is frequently specified *before applications* that use the subsystem *are known* [emphasis added], the designer may be tempted to "help" the users by taking on more function than necessary. Awareness of end-to-end arguments can help to reduce such temptations." (Saltzer, Reed, & Clark, 1984)

Acceptance of and adherence to this principle, also expressed by its corollary "innovations must be added at the edge and on top of the 'stupid network'" (Isenberg, 1997), should drive the development of a "communication subsystem" for interoperable digital mobility-as-a-service applications. The IF has been conceived and implemented as such a "communications subsystem", clearly identifying what functions should *not* be placed in the network or applications, respectively:

- Applications should *not* be concerned with the mechanics of interoperability, e.g. locating, accessing, moving, mapping resources needed in the distributed computation.
- The IF should *not* be concerned with business requirements and processes, or regulatory constraints and provisions.

In rest of the paper, after discussing some related works that are relevant to the concept of the IF (Section 2), we introduce the conceptual architecture of the IF (Section 3) and provide some details about its main components and their deployment strategies (Section 4). We then briefly outline the proof-of-concept implementation of the IF (Section 5), before concluding in Section 6.

## 2. Related works

In this section we overview solutions for tackling the interoperability problem particularly in geographically and administratively distributed systems characterized by heterogeneous actors and standards. The analysis focuses on the study and evaluation of various software architectures; it allowed us to identify the pros and cons of different architectural styles in practice, which in turn helped us design our system in such a way to address the interoperability requirement though the most suitable approaches.

Similar to the transportation sector, Internet of Things (IoT) is a heterogeneous domain composed of millions of devices interacting with each other through heterogenous standards, communication protocols, and vendor-specific APIs. IoT ecosystems can be spread over vast geographical boundaries and comprise a wide range of stakeholders. In this domain, we can cite the work carried out in the bIoTope project toward the construction of a federated IoT system (The bIoTope Project , n.d.). The bIoTope project follows a top-down interoperability strategy, where interoperable communications and interactions are achieved by encouraging the community to use certain (open) standards. From an architectural point of view (Kolbe, Kubler, Robert, Le Traon, & Zaslavsky, 2017), the project has employed a federated architecture where the service and data providers register their services/data in a distributed repository. The unified interpretation of the messages/data communicated between the actors of the system is guaranteed by the strict assumption that all would use a particular set of standards. In other words, the semantic interoperability is hardcoded in the system, which hampers loosely coupled and technology-independent interoperability. OpenIoT is another IoT middleware infrastructure aimed to support flexible collection and filtering of information streams and make them available for generating and processing business/applications events (OpenIoT, n.d.). To achieve full interoperability, OpenIoT has defined an ontology as an extension of Semantic Sensor Network (SSN) (Compton, et al., 2012) that is one of the best-known ontologies in the IoT domain developed by the World Wide Web Consortium (W3C). From the architectural point of view, OpenIoT is a middleware-based approach enabling the semantic unification of diverse IoT applications in the cloud. The raw sensed data that are collected from physical and virtual sensors are cast to the cloud layer (Linked Stream Middleware, LSM) and eventually converted to semantically-annotated data in RDF format, where users could initiate discovery request over data. In addition, the cloud layer stores metadata regarding the sensors and their functions. LSM then exposes SPARQL endpoints to enable the exploration of these semantically-annotated data.

Another computing domain where interoperability has become a major challenge is cloud computing. Similar to our case, cloud computing is composed of distributed resources (physical, networking and services) working together to realize the promise of cloud computing, that is a global market of collaborative services (Loutas, Kamateri, Bosi, & Tarabanis, 2011 ). The interoperability problem in the cloud stems from the intense competition among the giant cloud providers such as Google, Amazon and Salesforce, which makes them reluctant to converge towards unified standards. Cloud interoperability, in general, is "the ability to write code that works with more than one Cloud provider simultaneously, regardless of the differences between the providers" and semantic interoperability is concerned with how different cloud systems express and understand the same information (Loutas, Kamateri, Bosi, & Tarabanis, 2011 ). The work presented in (Loutas, Kamateri, Bosi, & Tarabanis, 2011 ) proposed an open Reference Architecture for Semantically Interoperable Clouds (RASIC). RASIC acts as a mediator between cloud providers with the aim to resolve semantic conflicts. However, their main contribution is the development of a model for a generic API through which cloud consumers can specify their requirements. The authors discuss how current cloud providers are offering analogous services with similar actions and properties, but through heterogenous APIs, names and structures. They suggest that ontologies and a unified modeling approach could be employed to overcome such interoperability issues. In particular, the authors have developed an ontology and vocabulary for the semantic annotation of both services and resources in the cloud; then, through their semantic engine, a cloud provider could formally (using RDF and OWL) add the mapping between its data model and the reference architecture's models. Using these mappings and suitable reasoning approaches RASIC could find semantically-matched concepts and resolve any semantic conflicts at runtime. In this context, (Rezaei, Chiew, Lee, & Aliee, 2014) proposes a semantic interoperability framework for Software as a Service (SaaS) systems in cloud computing environments The authors discuss that while syntactic interoperability has been achieved in the cloud, it severely lacks semantic interoperability. To overcome this issue, they proposed a broker-based approach which stands between cloud consumers and providers and takes care of the operations needed to ensure interoperability. In addition, the broker plays a central role to provide semantic interoperability. The cloud providers submit the syntactic description of their services through WSDL specifications to the broker. The semantic interoperability layer of the broker then creates the semantic description of the service explaining in a unified manner what it does and what are the requirements, limitations and service quality.

To sum up, our study shows that modular and service-oriented approaches seemed to be the key to solving the interoperability problem. In addition, the survey helped us recognize different requirements and the critical aspects that must be considered for the development of an interoperability framework. For instance, there was a common agreement among all surveyed works on the necessity and significance of a semantic-based approach and of the development of a shared ontology to achieve semantic interoperability in addition to syntactic interoperability.

## 3. Interoperability Framework: Conceptual Architecture

Figure 1 represents the high-level conceptual architecture of the IF, which is one of the Technology Demonstrators (TD) of the fourth Innovation Program (IP4) of Shift2Rail, dealing with "IT Solutions for Attractive Railway Services". The IF enables interoperability of all multimodal services by relieving consumer applications from the task of locating, harmonizing and understanding multiple, heterogeneous and independent sources of data, events, and service resources, which are made available as web services. The provision of this semantically-consistent overall abstraction of the distributed nature of the required digital resources is supported by two underlying specific logical abstractions, namely Data Abstraction and Service Abstraction, as represented in Figure 1.

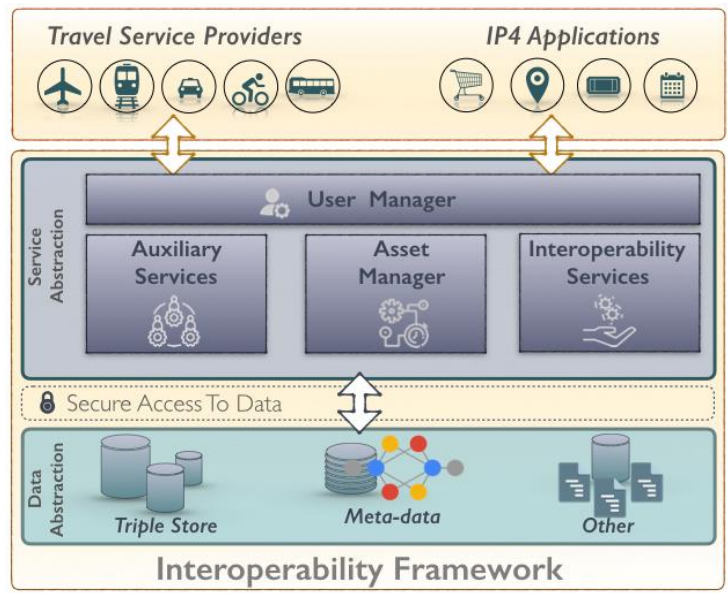

*Figure 1. Conceptual Architecture of the IF*

**Data Abstraction.** Starting at the bottom of Figure 1, "data" may comprise a wide range of categories and types such as transportation ontologies, supply chain and logistics data, code lists, ticketing and payment data, historical mobility data, traffic data, fares data, etc. In the absence of a unified standard, such data are modelled and represented following a heterogeneous set of specifications, vocabularies and data models, which in turn hinders smooth collaboration among different parties. One of the central goals of the IF is to overcome this fragmentation through the provision of semantic interoperability. Semantic interoperability in the IF is achieved by providing: (i) the association of machine-readable semantics, or 'domain knowledge', with data through the terms of a reference ontology; (ii) the automated conversion and enrichment of heterogeneous data into a common shared semantic graph. Data Abstraction covers the application of semantic interoperability to back-end databases and to the management of operations for collection, storage, and retrieval of shared data, as well as ontologies, vocabularies and meta-data provided both by external parties and by internal components of the IF in the Service Abstraction.

**Service Abstraction.** One of the core design goals of the IF's conceptual architecture is to envision a flexible, extensible and reusable infrastructure that could adapt and evolve gradually as both new technology and new requirements emerge. The Service Abstraction is designed, following a modular and service-oriented approach, as a set of self-contained, reusable and composable services. Services in the Service Abstraction are not exclusively limited to those initiated by the IF. On the contrary, to form a collaborative and open ecosystem, the IF facilitates any party in the incorporation of new services to the framework by supporting the necessary processes for the registration, discovery and deployment of such services. More specifically, services in the IF fall in three categories: Asset Manager, Interoperability Services and Auxiliary Services. While the services offered by the Asset Manager cover the generic operations required for overall management and accessibility of the IF, interoperability services include a set of special-purpose operations designed to mask the heterogeneous and distributed nature of transportation operators' ICT systems. Finally, any party could expose its services to be explored and utilized by others; in the scope of the IF the services supporting these features are called Auxiliary Services.

**4. Interoperability Framework Components: Design and Deployment**

At its core, the IF handles two types of entities, Assets and Users. In the scope of the IF, an *Asset* is an artifact that has some descriptions and it is discoverable by Users and other Assets. In other words, any resource that a generic actor of the transportation domain might be interested in – to read, share and use – has been referred to as an Asset in the IF. Figure 2 depicts three categories of Assets: Data, Utilities and Components, along with their subcategories (whose sets of instances can overlap). Category Data includes any types of data mentioned in Section 3, while Utilities and Components are the realization of the Service Abstraction also discussed in Section 3. In other words, *Utility Assets* (e.g., Ontology Editor, Mapping IDE) and *Component Assets* (e.g., Converter) are tools and services that foster interoperability. Notice that such Assets might be provided and used by external actors, as well as the IF itself. In addition, as shown in Figure 2, a *Component Asset* could be packaged as different deployable units; this enables multiple deployment and engagement choices for the IF's clients and widens the usability of the IF, as discussed in Section 4.1. Finally, the type of an Asset determines its lifecycle and the specification of its description.

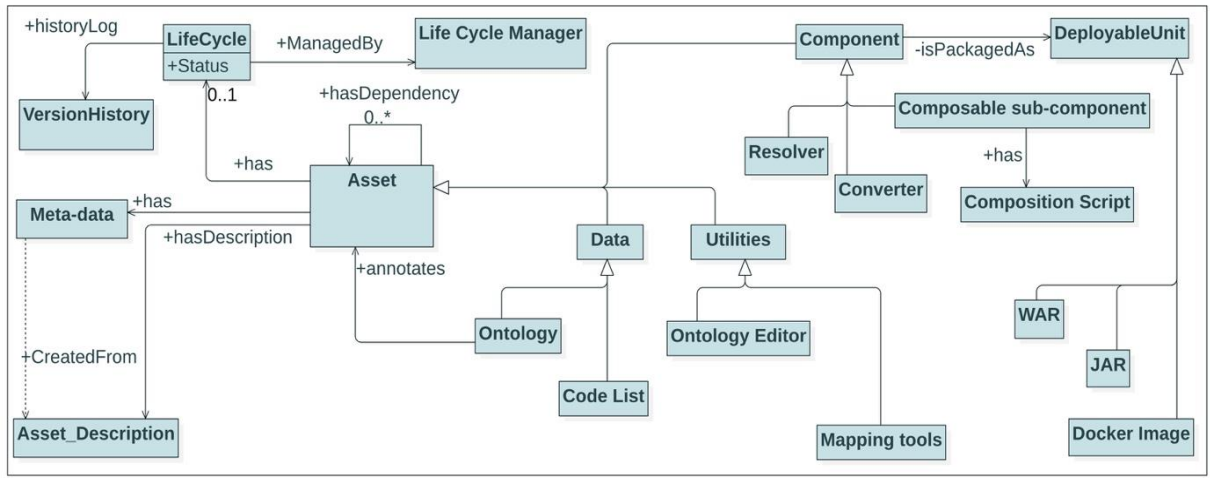

*Figure 2 Asset domain model*

The other type of entity at the core of the IF is the *User* (which is handled by the User Manager component shown in Figure 3). A human user of the IF might be an individual or a representative of a transport operator. In general, the possible interactions of *Users* with the IF could be encapsulated in two different logical roles, namely the IF providers and the IF consumers. The IF providers include a wide range of service/infrastructure providers in the transportation domain, including transport organizing authorities, transport service providers, infrastructure managers, retailers and travel agency distributors (SPRINT, 2019). Similarly, the IF consumers are also transportation actors such as travel service providers, social networks, and IT suppliers and software applications. The User Manger component governs the access rights and authorization for granting/denying various permissions to get access and operate with Assets based on the user's role and following a role-based access control mechanism.

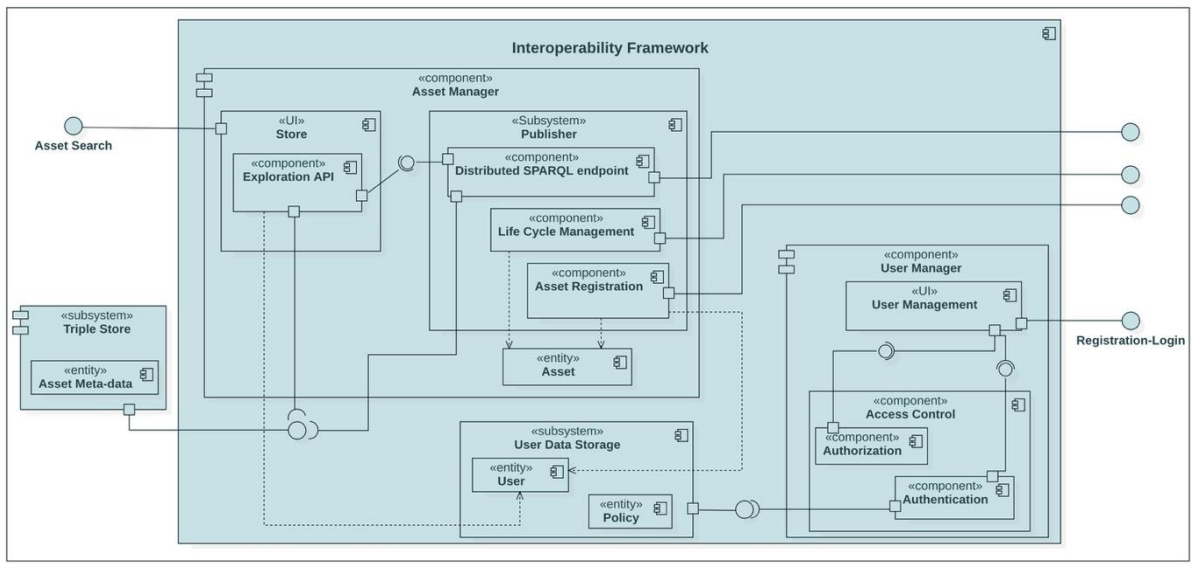

*Figure 3 IF component Diagram*

As depicted in Figure 3, **Asset Manager** is a pivotal component of the IF. It offers the basic functions to publish, share, discover, maintain and manage various artefacts that might be published/used by external and internal components of the IF. It acts as a catalogue of Assets which are subject to specific publication processes. The catalogue stores the asset's metadata in RDF according to the DCAT-AP (Goedertier, 2013) and ADMS (Dekkers, 2013) vocabularies. Its two main user interfaces are the Publisher and the Store, providing the possibility to contribute assets to the ecosystem and to look for specific assets in the catalogue. The Asset Store is the end-user interface, showing the available assets types and their respective assets. It is possible to configure it so to let users obtain just a preview of the asset information, forcing them to submit an access request to obtain the full visibility over the asset metadata, its attachments and specific functions. The Assets Publisher is used to contribute an asset to the catalogue, and to manage its lifecycle. Assets to be used in a wide ecosystem need to be managed in a consistent way to foster trust among the parties. To that extent, each asset type in the Assets Publisher is linked to a lifecycle process, which covers all the aspects from the initial submission to the final publication, plus all the aspects related to change management and the effects of a change on dependent assets. Since such lifecycle processes (expressed using the BPMN 2.0 standard) define roles and responsibilities in a distributed environment, they become the real "contract" between the parties participating in the ecosystem.

A Currently, the initial set of asset types defined to be hosted by Asset Manager is composed by *Ontologies*, *Datasets* (of Data Category), *Resolvers, Converters* (of Component Category), and *Ontology Editor and Mapping IDE* (of Utilities Category). An Exploration API asset type (of Component Category) has also been defined to facilitate access to a potentially high number of asset metadata descriptions via common Web technologies. This asset type is used to describe parametric SPARQL queries, which are then exposed by Asset Manager as Web APIs with a mechanism akin to grlc (Meroño-Peñuela & Hoekstra, 2016) and basil (Daga, Panziera, & Pedrinaci, 2015). Using a normal HTTP GET request, the user can provide values to the parameters of the SPARQL query and obtain the corresponding results. Publishing parametric queries as assets in the catalogue allows a much higher level of control over the users' behaviors, since users are only allowed to call specific Web APIs according to their security permissions.

The IF provides applications with a semantically consistent abstraction of resources that are available to perform a computation to create an integrated mobility solution for customers. However, for each specific such computation, a context must be established identifying the scope and the resources that are actually needed in the specific task. **Interoperability Services** are components of the IF exposed as services[‡] that are used by client applications for this purpose. The main Interoperability Services are Converters and Resolvers.

*Resolver* services provide access, location, relocation and replication transparency to interacting applications, masking them from the physical distribution, access protocols and formats of meta-data and data resources available in the Data Abstraction. A resolver could be seen as a special-purpose discovery component focused on the discovery and access of a specific resource. For instance, a Location Resolver returns a list of Stop Places within a requested radius from a point specified by its geographical coordinates; a Travel Expert Resolver returns a list of web services that create mobility offers for a specified pair of departure and destination Stop Places.

*Converters* address the interoperability problem through an automated transformation of heterogeneous standards to a formal, sharable and machine-readable semantic reference model and vice versa. Different parties can interact seamlessly while each one relies on its own native data specification. A Converter is a software artifact (either in the form of a service or a library) to be used to enable interoperability between systems communicating according to different specifications. A converter implements a data processing pipeline which *lifts* incoming message data into RDF according to a specific ontology, and then *lowers* the RDF triples into the specific destination format, as explained in (Carenini, et al., 2018).

*4.1. Deployment Strategies of Interoperability Components*

The heterogeneity of the IF's clients, which might have different requirements and application domains, led us to design the architecture and deployment of the IF so as to cover a wide range of use-cases. Currently, we are anticipating three ways for providers/clients to offer/use interoperability components, which are descried in the following.

---

[‡] IF policy is to encourages servitization of its components and interoperability functions, hence the primary packaging and publishing strategy is through a service. However, to address a wider range of use cases, other deployment approaches also are supported (see Section 4.1).

**Direct Access** is a standard approach for facilitating a loosely-coupled and service-oriented interoperability among transportation actors. A publisher can advertise an already-running component in the IF by providing a generic description for it, which includes its endpoint. Asset Manager then creates meta-data out of this description to make the component discoverable by the IF clients through distributed SPARQL endpoints. In this case, the IF is mainly a service repository that bridges the gap between service descriptions and semantic web technologies, thus fostering a wider discovery range. After the discovery phase, the role of the IF ends, and the client is redirected to the provider system, where it engages with the desired service.

**Runtime executable environment** is as an extension of the previous model that promotes a Platform as a Service (PaaS) approach. It broadens the role of Asset Manager beyond that of a catalogue manager. Asset Manager in this case becomes a command-and-control tool that actively manages deployable artifacts onto a cloud platform. To this end, either the service provider itself, or any other external infrastructure provider should authorize the IF to use such a cloud environment upon user request. Accordingly, after the discovery phase, Asset Manager takes the appropriate executable artefact of the selected component, runs it on the target cloud environment, and returns the endpoint that allows the client to access the service.

Finally, to cover use cases where clients prefer to run the desired component locally or to integrate it as an internal part of their system, the IF supports a **Direct Download** approach, where the publisher uploads the component implementation on the IF, which then lets users download it. More precisely, as discussed above and depicted in Figure 2, a component of the IF could be wrapped up and published trough different downloadable and runnable artefacts, including container images, JAR and WAR files. A component might be packaged in multiple forms, which gives the client the possibility of choosing the most suitable method to engage with the desired component. For example, if the client's internal system follows a micro-service-based architecture, which is most often realized through technologies such as Docker (Docker, n.d.), then a docker image of a component (e.g., Converter) that could readily be deployed on a Kubernetes cluster (Kubernetes, n.d.) and integrated with rest of the system seems the best option for the user. In other cases, if the client prefers a conventional service-oriented approach, a component packaged as WAR file that could be integrated in its internal environment is a suitable option.

**5. Proof-of-concept Implementation**

This section briefly describes a pair of key elements of the IF that have been developed within the SPRINT project. In particular, it first presents the current implementation of the Converter component (called Chimera), which is based on and extends the technology first developed within the ST4RT project (Carenini, et al., 2018); then, it describes the main features of the Asset Manager.

*5.1. Chimera*

A conversion process can be interpreted as a data pipeline composed of several blocks, each one devoted to a specific task. Such tasks can be: (i) integrating the incoming message with other data, (ii) transforming the message content, and (iii) reformatting the transformed message into the desired output format. In the context of RDF-based conversion, these blocks become: enriching the "conversion context" (the data pool used as a basis for the transformation) with RDF datasets and ontologies, performing *lifting* according to Java annotations or RML mappings, and performing *lowering* to obtain the desired output.

Chimera is the SPRINT reference implementation of the Converter. It is built upon Apache Camel as a framework to let developers create custom converters by choosing different input and output communication channels and protocols, and by integrating RDF datasets and ontologies in the right order before performing the lowering.

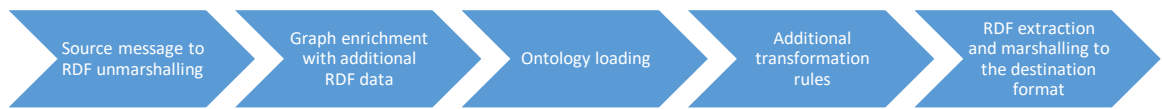

*Figure 4 RDF-based conversion process using Chimera*

Using the Enterprise Integration Patterns (EIP) implemented by Camel, the conversion process can be implemented by a pipeline of "enrichers", as depicted in Figure 4. At the beginning of the conversion an RDF graph is created, which is then used to hold all the RDF triples produced by each processing block. At the end of the pipeline, the lowering block extracts data from the context provided by the RDF graph and creates the output message.

Different technological alternatives are provided for each conversion pipeline block. To perform the lifting, developers can choose to use either the declarative approach provided by the RML lifting block, or the annotations-based block developed during the ST4RT project and based on the Empire framework. To perform the lowering, two building blocks are available, one again based on Java annotations developed in the ST4RT project, while the other is based on SPARQL for data extraction and on the Velocity template engine for the actual composition of the output message.

*5.2. Asset Manager*

The Asset Manager, as explained before, is the service devoted to supporting the IF ecosystem day-by-day operations related to data sharing. Its two main tasks are allowing the publication of asset descriptions and coordinating actors related to different transport operators to ensure that each asset description has been published by the right actor in accordance to the rules shared by ecosystem participants.

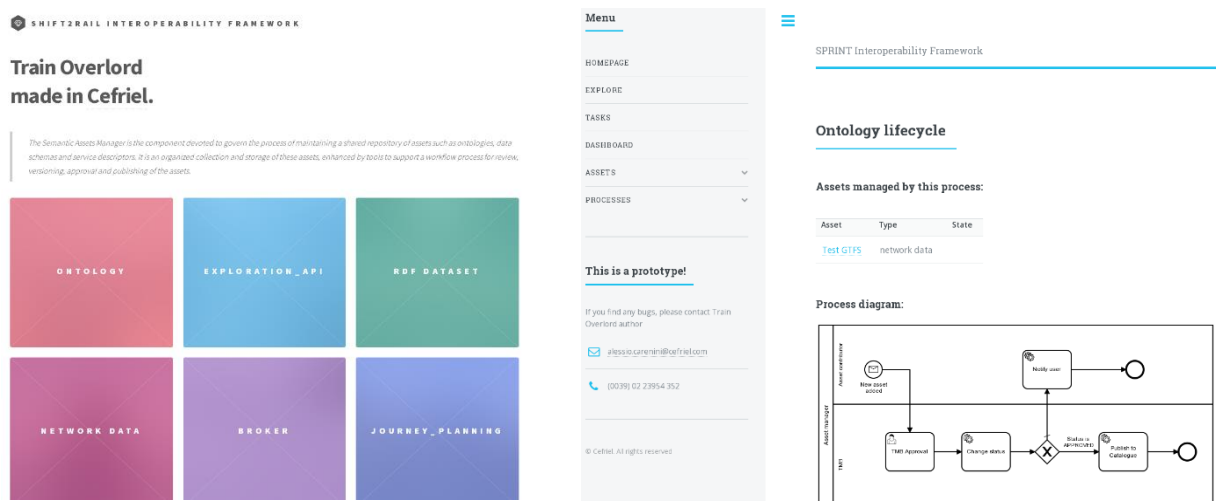

*Figure 5 Asset manager Store and Publisher*

Internally, each of the main features is implemented by a separate server component. The Catalogue provides two separate user interfaces for publishing and accessing data provided by transport operators. This decoupling allows designing a technical-oriented user interface devoted to publishing, and a more store-oriented user interface presenting the published information in an easier way. We chose to implement the Catalogue and its related features (such as user registration, authorization and a REST interface) through the Django framework (Django, n.d.). The Catalogue is also able to convert the asset descriptions in RDF according to a specific ontology which extends DCAT-AP and ADMS, and to feed the resulting triples to an RDF repository (Ontotext GraphDB (GraphDB, n.d.)). This allows keeping a well-defined representation of the metadata of each asset, and it enables using a SPARQL endpoint to query it.

The Lifecycle manager is instead devoted, as introduced before, to the governance of the publication of asset descriptions. Governing the publication process means coordinating both humans and services. As an example, humans need to be aware that a new asset description has been published, and that it needs a formal approval before further actions can be carried out. According to human decisions, services implementing specific features can be called, such as converting the asset description in its RDF representation according to a set of ontologies and vocabularies. Since lifecycle management processes can be arbitrarily complex, we chose BPMN 2.0 as a way to specify them, as it allows extending the XML representation of "Service Task" and "Human Task" to allow a vendor to be able to actually execute a process. We then chose the open source Camunda process engine (Camunda, n.d.) to support BPMN execution, and we integrated it with the Catalogue so that whenever an asset description is added or modified, an event is sent to Camunda via its Web API to trigger the execution of the right process.

Having a catalogue of assets, described according to an ontology, enables the possibility to enrich the contents of the catalogue itself by automatically executing a set of pre-defined tasks coordinated by the lifecycle management.

Adding an ontology can start an automatic documentation job which is able to add both its textual and visual representation to the ontology asset. Describing the contents of a Converter, in terms of expected source and destination specifications, can drive the actual synthesis of a Converter artifact which is then provided to the user. Such automation jobs are quite common in Continuous Integration and Delivery, and we chose to integrate the open source Jenkins server (Jenkins, n.d.) as one of the internal components of the Asset Manager.

## 6. Conclusions

This paper presents the SPRINT approach for the design and development of the reference architecture of the IF, which is a central element in the realization of the Shift2Rail objective of a complete transformation of the European Transportation ecosystem into a global asset marketplace, liberated from technological barriers. Specifically, the proposed reference architecture realizes the objectives of the IF in two ways. Firstly, through the Asset Manager, which masks the complexity of interoperability to travel applications by publishing uniform abstractions of services, and which enables travel applications to communicate among them uniformly (e.g., web service/API interfaces, communication protocols). Secondly, by providing additional technical means to automate and facilitate seamless and smooth cooperation of heterogeneous and fragmented transportation actors and to enable them to operate on the web of transportation data.

**References**

*Camunda*. (n.d.). Retrieved from https://camunda.com

Carenini, A., Dell'Arciprete, U., Gogos, S., Pourhashem Kallehbasti, M. M., Rossi, M., & Santoro, R. (2018). ST4RT-- Semantic Transformations for Rail Transportation. *Transport Research Arena TRA 2018*, (p. 1-10).

Compton, M., Barnaghi, P., Bermudez, L., Garcia-Castro, R., Corcho, O., Cox, S., . . . others. (2012). The SSN ontology of the W3C semantic sensor network incubator group. *Web semantics: science, services and agents on the World Wide Web*, 25--32.

Daga, E., Panziera, L., & Pedrinaci, C. (2015). Basil: A cloud platform for sharing and reusing SPARQL queries as Web APIs. *CEUR Workshop Proceedings*, *1486*.

Dekkers, M. (2013). Asset description metadata schema (adms). *W3C Working Group*.

*Django*. (n.d.). Retrieved from https://www.djangoproject.com

*Docker*. (n.d.). Retrieved from https://www.docker.com

Goedertier, S. (2013). DCAT application profile for data portals in Europe.

*GraphDB*. (n.d.). Retrieved from http://graphdb.ontotext.com

Isenberg, D. (1997). The rise of the stupid network. *Computer Telephony 5.8*, 16-26.

ISO/IEC 10746-3:2009. (2009). *Information technology Open distributed processing Reference model: Architecture.* International Organization for Standardization.

*Jenkins*. (n.d.). Retrieved from https://jenkins.io

Kolbe, N., Kubler, S., Robert, J., Le Traon, Y., & Zaslavsky, A. (2017). Towards semantic interoperability in an open IoT ecosystem for connected vehicle services. In *2017 Global Internet of Things Summit (GIoTS).* IEEE.

*Kubernetes*. (n.d.). Retrieved from https://kubernetes.io

Lay, G. (2014). *Servitization in industry.* Springer.

Loutas, N., Kamateri, E., Bosi, F., & Tarabanis, K. (2011 ). Cloud computing interoperability: the state of play. *IEEE Third International Conference on Cloud Computing Technology and Science.*

Meroño-Peñuela, A., & Hoekstra, R. (2016). grlc makes GitHub taste like linked data APIs. *European Semantic Web Conference*, (p. 342-353).

*OpenIoT*. (n.d.). Retrieved from http://www.openiot.eu

Rezaei, R., Chiew, T. K., Lee, S. P., & Aliee, Z. S. (2014). A semantic interoperability framework for software as a service systems in cloud computing environments. *Expert Systems with Applications, 5751--5770*(Elsevier).

Saltzer, J. H., Reed, D. P., & Clark, D. D. (1984). *End-to-end arguments in system design.* echnology 100 .

SPRINT. (2019). *D3.2 - PERFORMANCE AND SCALABILITY REQUIREMENTS FOR THE IF.* Tratto da http://sprint-transport.eu/Page.aspx?CAT=DELIVERABLES&IdPage=1e2645be-e780-4d99-8117-bae57b67b453

The bIoTope Project . (s.d.). *bIoTope*. Tratto da https://biotope-project.eu/

Vandermerwe, S., & Rada, J. (1998). Servitization of business: adding value by adding services. *European management journal 6, no. 4*, 314-324.